\documentclass[prb,twocolumn,showpacs,preprintnumbers,amsmath,amssymb]{revtex4}

\usepackage{amsfonts}
\usepackage{latexsym}
\usepackage{graphicx}

\newcommand{\be}{\begin{equation}}
\newcommand{\ee}{\end{equation}}

\begin{document}

\title{Quantization of conductivity of nanotechnological point contact.
Simple derivation of the Landauer formula}

\author{T. Mishonov}
\email[E-mail: ]{mishonov@phys.uni-sofia.bg}
\author{M. Stoev}
\email[E-mail: ]{martin_stoev@abv.bg}
\affiliation{Department of Theoretical Physics, Faculty of Physics,\\
University of Sofia St.~Kliment Ohridski,\\
5 J. Bourchier Boulevard, BG-1164 Sofia, Bulgaria}

\date{\today}

\begin{abstract}
A simple methodical derivation for Landauer quantization of the
conductivity is derived as a simple consequence of the Bohr
quantization laws. The level of explanation corresponds to
high-school level of Physics education and can be used as popular
lecture for students. The purpose of the work is to introduce
students in an achievement of nano-technology which is relevant to
the future electronics. Using only the fundamental laws of
quantization students can understand a contemporary experimental
research and to follow future development in the field.
\end{abstract}


\preprint{Scientific-Methodical Journal of Physics \textbf{30}(\textbf{48}), b. 1,  16-22 (2005) (Sofia, in Bulgarian) http://www.physica.hit.bg}
\maketitle

\section{Introduction}

The permanent miniaturization in the electronic technology leads to
the need of mentioning the quantum effects, while exploring the
electron transfer in nano-structures. For example in the common
copper cables the electrons are moving the same way as in a bulk
poly-crystalline metal, but when the wideness of the conductor is in
the sub-micron area, there are observed quantum effects. In these
extremely thin wires, or as they are usually called nano-wires, the
electron moves by the length of the conductor like a wave; the
wideness of the wire must be less than 100 nm. We must calculate the
current, caused by the different types of electron waves, which are
conveyed through the nano-wire the same way as the television signal
is transferred through the coaxial cable. When the nano-wires are
small enough and are very precisely made, the electron diffusion is
low enough and we can think of them like ``flying'' through the
whole wire with a constant velocity $v$ and momentum $p.$ This kind
of movement is called ballistic, like a free flying bullet. The
problem of calculating the electric conductivity of a nano-wire,
connecting two bulk conductors as is shown on Fig.~1, was solved
before nearly half a century by Rolf Landauer, who worked then in
IBM.
\begin{figure}[t]
\centering
\includegraphics[width=0.6\columnwidth]{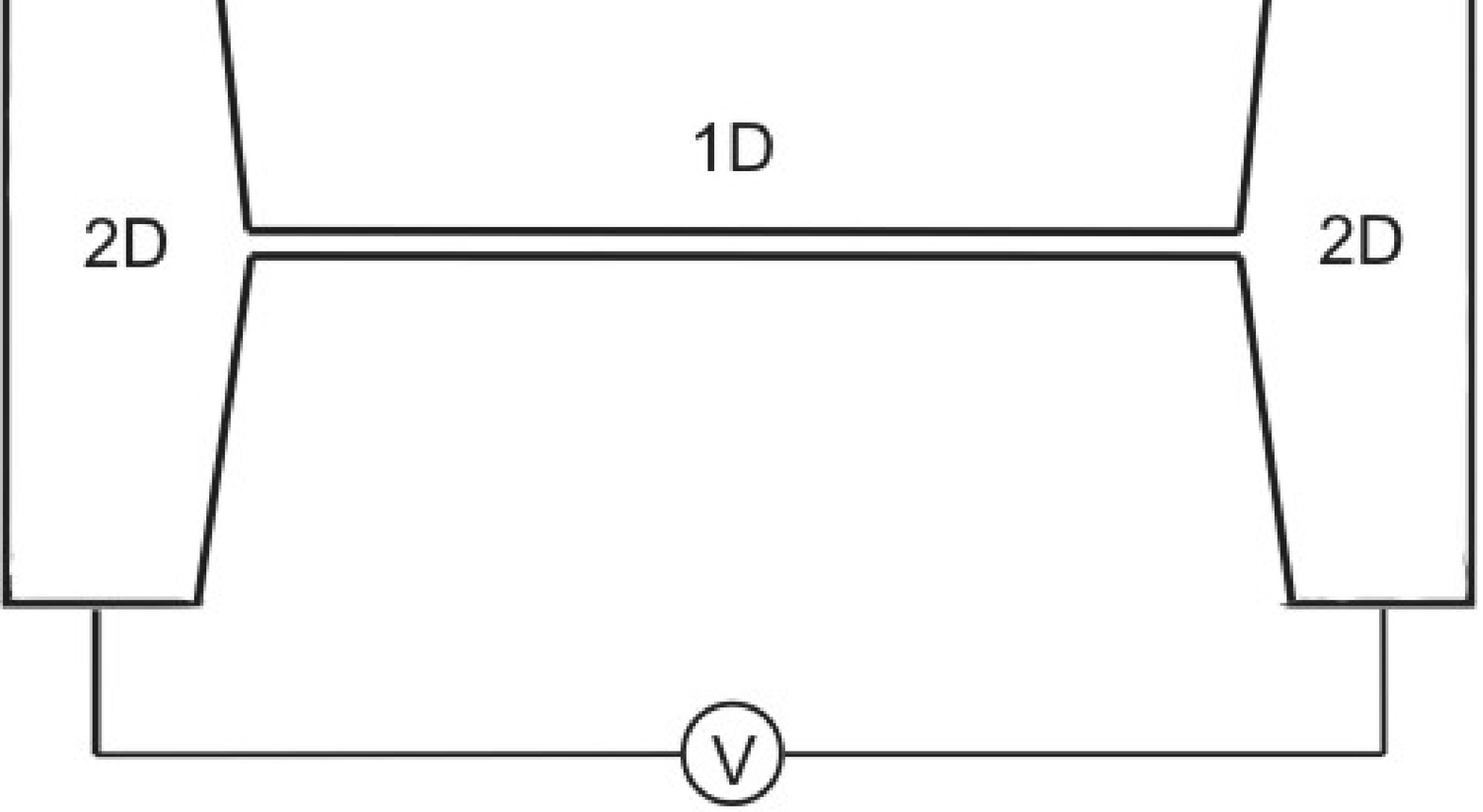}
\caption{One dimensional (1D) channel connecting two bulk
two-dimensional (2D) conductors. Applied voltage creates current
and the conductivity is quantized.
\label{fig:1}}
\end{figure}
The predicted by Landauer quantization of the conductivity was
demonstrated very precisely by a split gate field effect
transistor\cite{Kouwenhoven,Wees:88,Wees:91}, shown schematically
on Fig.~2.
\begin{figure}[t]
\centering
\includegraphics[width=0.6\columnwidth]{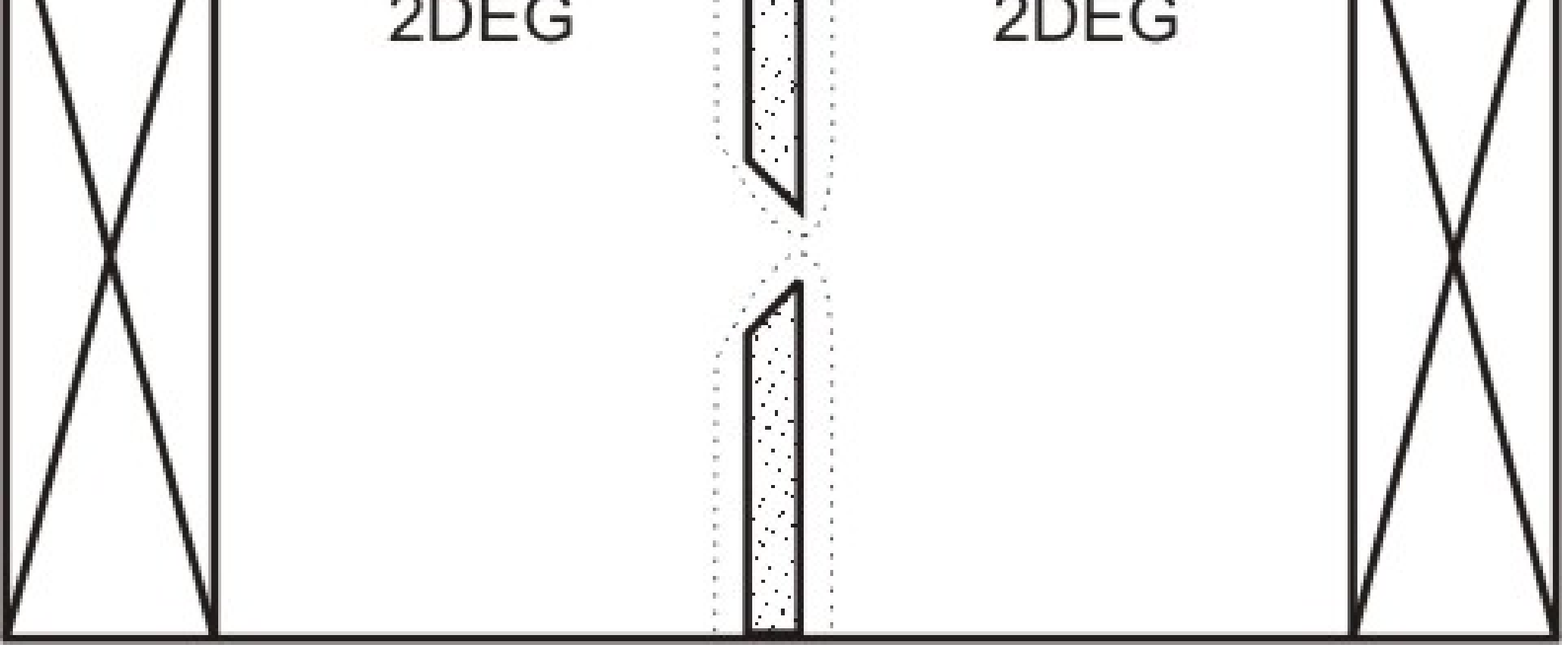}
\caption{Realization of 1D conductor (point contact) by split-gate
technology field effect transistor. Gold split-gate rejects the
electrons from the two dimensional electron gas (2DEG) in thin GaAs
layer. The levels $\mu_1$ and $\mu_2$ of the two Fermi seas are
different.
\label{fig:2}}
\end{figure}
This theory is used for the analysis of the work of
sub-micron nano-structures, as the whole contemporary electronics
are based on the nano-technology. In leading electronic companies
there were made fundamental explorations of the properties of
nano-structures in different conditions; for example, there was
examined their behavior in very low temperatures. The purpose of
the present work is to describe the experiments
\cite{Kouwenhoven,Wees:88,Wees:91} and to make a simple deviation
of the Landauer's formula, using only the fundamental principles
of Physics, formulated by Niels Bohr. For understanding the
deviation is necessary only familiarity with the atomic model of
Bohr, which is already taught in the high-school level of
education. That is why our work may be used by teachers, who want
to show their classes a new material from the contemporary
Physics, but some pupils may understand the present work themselves.
Usually while analyzing some occurrences the action of the border
is too low to be assumed. For example the heat capacity of a ring
and a cylinder made from the same material is practically
equivalent. That is why it's understandable that we will use the
conclusion of  a helping problem to analyze the conductivity of a
nano-wire. This helping solution concerns electrons, circulating
on a nano-ring. Instead of exploring a flow of electrons,
transferred in a nano-wire from the negative electrode. We will
concentrate on electrons, winding in one direction on a sub-micron
conductive ring with radius $r$. After the analysis of the
reference between the current and the potential we will ``cut''
the ring and will apply the results for the analysis of the
conductivity of nano-technology point contacts.

\section{Realization of the experiment}
Let us describe shortly the realization of the one-dimensional
electron conductor by semi-conductive nano-structures. Decreasing
the dimensionality is achieved in two separate stages, and
technologically the most important step is the establishing of
two-dimensional electron gas 2DEG. For that reason there are used
laminar semi-conductive structures. A thin coat of GaAs is placed
in AlGaAs. Closely to that coat the bulk AlGaAs is alloyed in a
very thin area with Si. The electron leaves the Si donors and fall
into the potential hole of GaAs. In this situation their
wave-functions are just standing waves, like the tremble of a
violin string. But in the flat of the interface the electrons move
like free two-dimensional particles and their wave-functions are
flat waves. In different words, we can assume that in the thin
coat the electrons are soaring like seagulls above the surface of
the sea. Their dissipation on the distant ionized Si donors is low
and the conductivity of the 2DEG is very high. The density of this
depraved electron gas is very high, and we have a two-dimensional
metal when the temperature is low. This metal is very well
insulated from the semi-conductor, in which it's situated. For
that structure the Pauli prohibition is valid and the electrons
fill all the electron states to some highest possible energy
$E_\mathrm{max},$ called Fermi energy, as it is schematically
depicted on Fig.~3.
\begin{figure}[t]
\centering
\includegraphics[width=0.6\columnwidth]{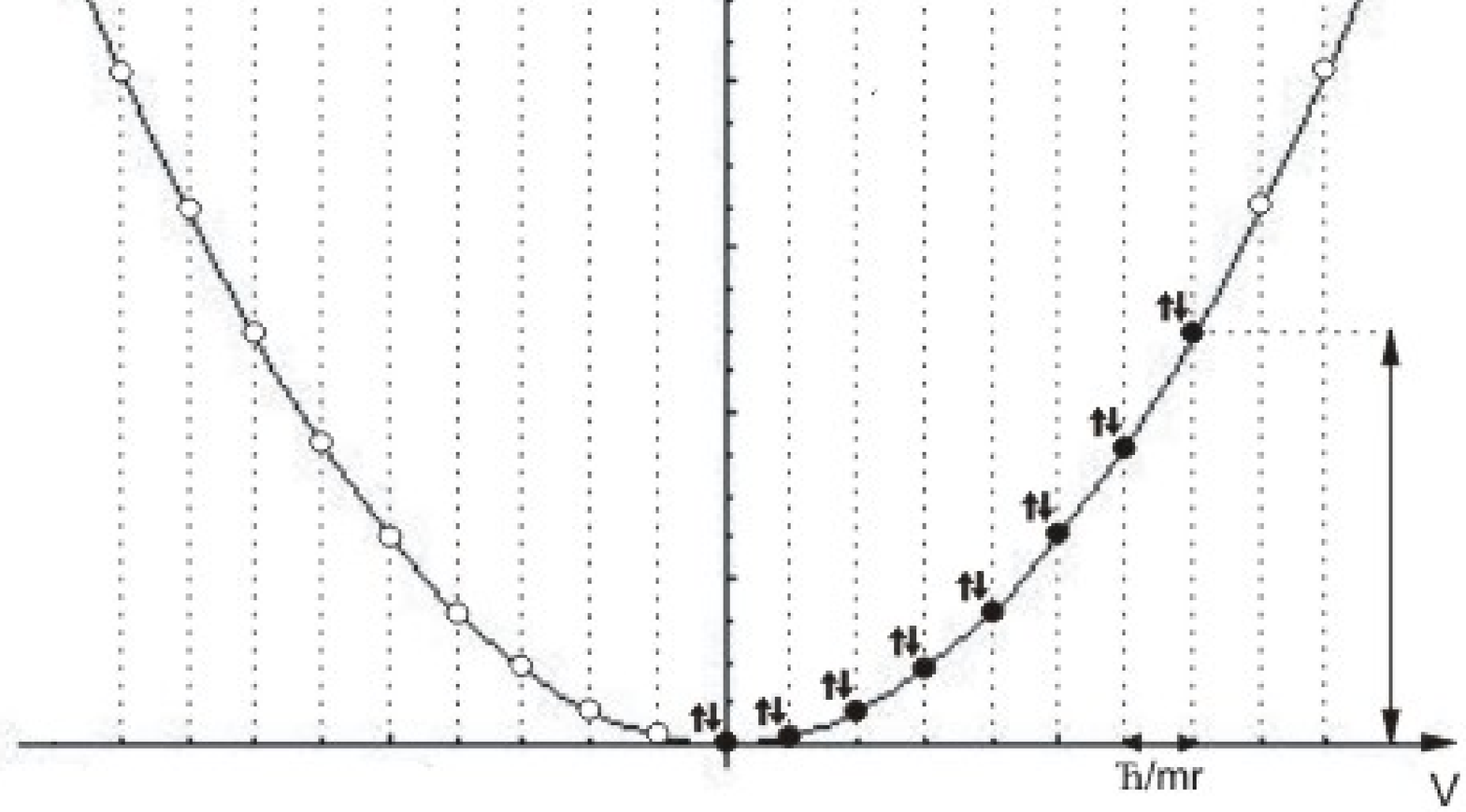}
\caption{Energy quantization when electrons circulate around a
ring with radius $r.$ Due to Bohr laws the angular electron
momentum is quantized. This leads to velocity and energy
quantization. Open circles presents empty electron orbitals.
Filled circles denote double electron occupied states. The
electrons moving in right direction have a maximal energy.
\label{fig:3}}
\end{figure}
The mentioned above filling of
electron position is analogous  to the filling with water of all
the volumes in Earth, beneath the sea level and that is why the phrase
Fermi sea is usually used. In our case the sea is two-dimensional.
The technological problem is to create one-dimensional channel
between two seas of that kind. For this purpose one split-gate of
two gold electrodes is used, evaporated on a semi-conductor
hetero-structure. That is how we create a field effect transistor,
demonstrated on Fig.~2. The source and the stock have an Ohm
connection with the 2DEG. When we put a strongly negative
potential on the gate, the electrons below it disappears, ejected
awry by the Coulomb repulsive force of the gate. That is how the
connection between the two sees is interrupted and the transistor
is bung and stopped. When we decrease the force of rejection by
changing the gate voltage the contact between the Fermi seas is
recovered and the electrons run from one sea, to the other through
the narrow one-dimensional channel, created under the split-gate.
In this phase of their transfer, the electrons have almost
one-dimensional motion - like waves in a waveguide. They do not
diffuse, but fly like bullets. That is why we call that condition
a Ballistic regime.
\begin{figure}[t]
\centering
\includegraphics[width=0.6\columnwidth]{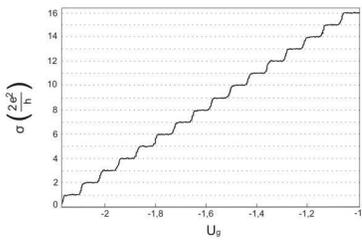}
\caption{Conductivity $\sigma$ of the field effect transistor
versus gate voltage $U_g$. Increasing the voltage opens new
electron channels. Every open electron channel gives one
conductivity quantum $2e^2/h$ to the whole conductivity which
creates the height of the steps.
\label{fig:4}}
\end{figure}
The wave-function of the electrons are flat
waves stretched by the length of the channel, and in the
perpendicular direction they are standing waves. The filling of
the electron states depends on the potential, so the gate voltage
determines the number of the open electron channels. Each of these
channels gives one quantum of conductivity to the whole
conductivity of the nano-technology point contact. Now we can
understand the step-looking nature of the relation between the
conductivity and the gate voltage in the low-temperature range,
shown on Fig.~4.Each step means that one more one-dimensional
channel has been opened between the Fermi seas. Increasing the
temperature means that the steps become smoothed-out and
indistinct, see Fig.~5;freezing brings us more step-looking
figure.
\begin{figure}[t]
\centering
\includegraphics[width=0.6\columnwidth]{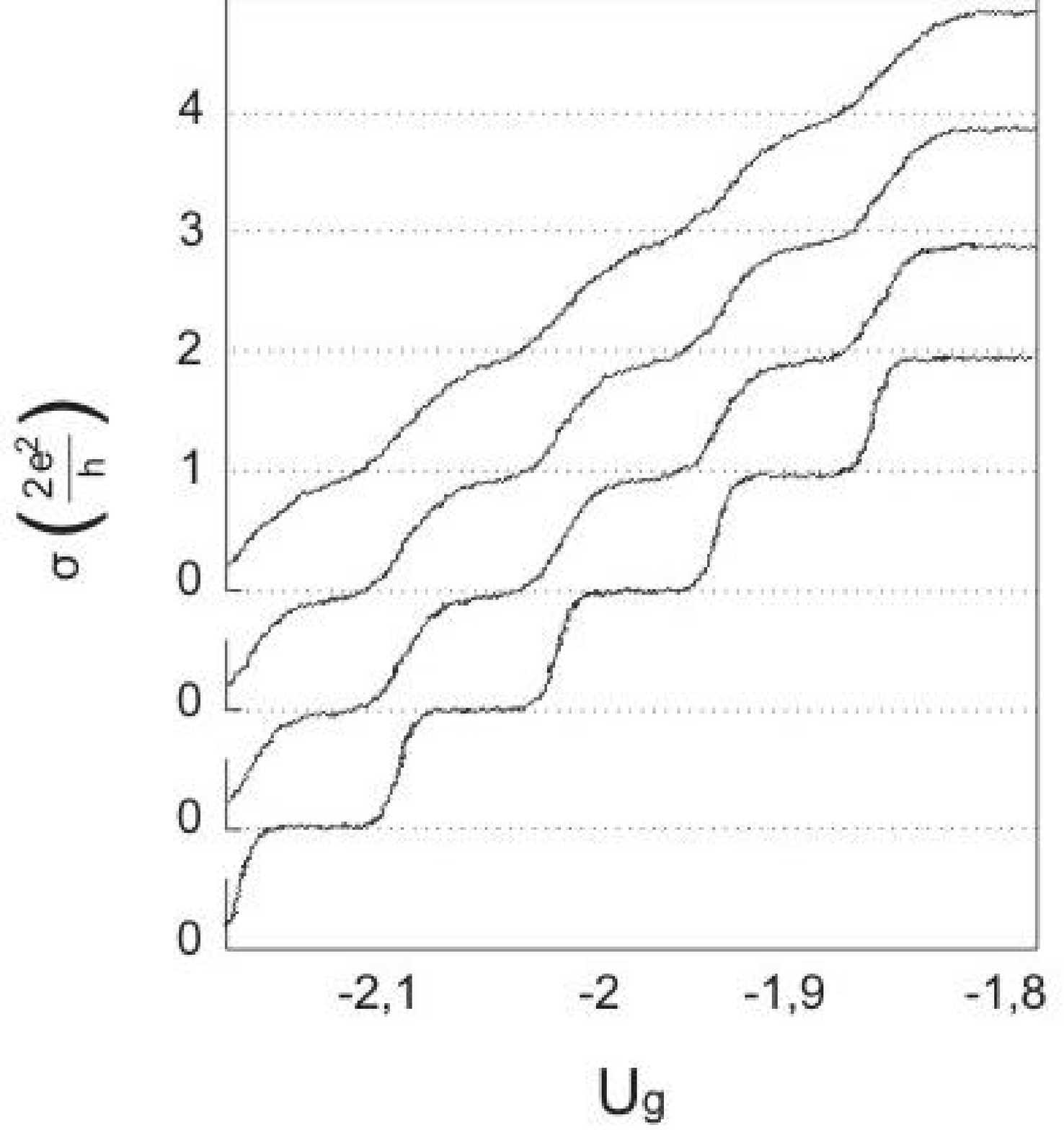}
\caption{Conductivity $\sigma$ versus gate voltage $U_g$ at
different temperatures. Increasing the temperature smears the
steps of conductivity quantization. At high temperatures the
conductivity quantization disappear.
\label{fig:5}}
\end{figure}
The height of the step may be computed with elementary
knowledges. These calculations were realized in the next part of
this work. If we want to be more accurate we have to take into
account currents in the opposite directions, when $U$ is a small
difference between the levels of the Fermi seas, but the result
remains the same. That is all we will say about the realization of
the experiment. In the next section the Landauer formula will be
deviated, describing the quantization of the conductivity. After
that we will describe the influence of the different temperatures
to the conductivity of the nano-technology point contact.

\section{Deviation of the Landauer Formula}
As we had already mentioned, we will analyze a mental experiment
with electrons, winding on a circle with radius $r$, like the
atomic model of Bohr. That experiment will help us to solve our
problem. When an electron circulates with velocity $v$ the period of its
round is $2\pi r/v$ (time is equal to the path divided by
velocity). And the average current $I_n$ is simply the electron
charge $e$ divided by the period (by definition the current is the
charge per unit time)
\begin{equation}
\label{currentE} I_n=\frac{ev_n}{2\pi r}.
\end{equation}
The index n means that we respect the quantum properties of the
electron. The integer $n$ shows us how many wavelengths of the
electron $\lambda_e=2\pi\hbar/p$ can be put together in perimeter
of the circle $2\pi r=n \lambda_e$. According to the Bohr's law
the angular momentum is quantized
\be mvr=n\hbar, \ee
where $\hbar=1.055\times 10^{-34}\,\mathrm{Js}$ is the Plank's
constant and the integer $n$ is called magnet quantum number in the
atomic physics, $m$ is the effective mass of the electron in the
crystal grid. From Bohr's law we can calculate the velocity:
\be v_{n}=\frac{\hbar}{mr}n .\ee
And we can put the results back in the formula, applied to the
formula for the current Eq.~(\ref{currentE}).
\be I_n=\frac{e\hbar}{2\pi mr^2}n. \label{1 electron currentE} \ee
The quantization of the velocity causes quantization of the
kinetic energy too
\be E_n=\frac{1}{2}mv_n^2=\frac{1}{2}\frac{\hbar^2}{mr^2}n^2. \ee
This formula explains the energetic spectrum of an electron,
winding on a circle with a fixed radius, graphically shown on
Fig.~3. Its now easy to calculate the whole current, caused by all
the electrons, circulating in one direction $v>0,$ assuming tat
all the electron states with energy in the range from $E=0$ to
$E=E_\mathrm{max}$ are filled with exactly 2 electrons each (with
spin ``up'' and ``down''). According to the Pauli's principle of
prohibition each electron state can be filled only with one
electron with certain spin and from the equation about the spectrum its obvious,
that the maximal energy $E_\mathrm{max}$ can be parameterized with
one big integer $N$ or with an item, having the character of
electric potential $U$
\be E_\mathrm{max}=\frac{1}{2} \frac{\hbar^2}{mr^2} N^2 =
eU.\label{FermiE}\ee
The whole current is equal to the sum of all the electron currents
\be I= 2 \sum^{N}_{n=0} I_n.\label{total currentE} \ee
In other words, a sum through all the electron states ought to be
made. We can use the formula for the arithmetic progression for
big enough numbers $N\gg1$
\be \sum_{n=1}^N n=1+2+3+\dots+N=\frac{N(N+1)}{2}\approx
\frac{N^2}{2}\gg1. \ee
We apply this formula for to the calculation of the currents
Eq.~(\ref{total currentE}), expressed by Eq.~(\ref{1 electron
currentE})
\be
I= 2 \sum^{N}_{n=0} I_n
 = 2\sum^{N}_{n=0}\frac{e\hbar}{2\pi mr^2}n
 = \frac{2e\hbar}{2\pi mr^2} \frac{N(N+1)}{2}.
\ee
The integer $N$ is much bigger than 1, so in a good approximation
we may assume that $N(1+1/N)\approx N$ and there we have
\be I= 2\frac{e}{2\pi\hbar} \left(\frac{\hbar^2N^2}{2mr^2}\right).
\ee
Here the expression in the brackets is the highest possible energy
of the electrons mentioned in Eq.~(\ref{FermiE}). Now we have the
opportunity to express the current by an electric potential
\be I=\frac{2e^2}{2\pi\hbar}U=\frac{2e^2}{h}U=\sigma U, \ee
following the tradition we have used the old Plank constant
$h=2\pi\hbar.$ By definition the conductivity $\sigma$ is the
current, divided to the voltage $\sigma=I/U$. In this whole
calculation we assume that the regime of the electrons,
transferred through the ring is ballistic, where the electron
diffusion is neglected. That is why the electron wavelength have
to be shorter than the average free path of the electron. The
result we achieved is applicable for short enough nano-wires,
including point-contacts too, where the one-dimensional movement
is in a very slight area. Now we have the Landauer formula,
describing the quantization of the conductivity of one-dimensional
conductor.
\be
\sigma_0=\frac{2e^2}{h}
 = \frac{1}{12906\;\Omega}
 = 77.5\,\mu\mathrm{Sm}.
\ee
In the realistic nano-technology point-contacts the conductivity
is achieved as a result of a great number of such one-dimensional
channels and to calculate the whole conductivity, we ought to
multiply the conductivity quantum $\sigma_0$ by the number of the
opened one-dimensional electron channels $K$
\be \sigma = K\sigma_0= K \frac{2e^2}{h}. \ee

\section{The influence of finite temperature}
Lets research the results, achieved by realizing the experiment.
In this work there was used the transistor, shown on Fig.~2. The
temperature, in which the measurements were made is close to the
absolute zero $T = 0.6$~K. Measuring the voltage, the conductivity
can be calculated for different values of the negative gate
potential $U_g$, applied to the point contact. A simple parallel can
be made, concerning the negative voltage, which can help the
reader to understand the graphical relation between $\sigma$ and
$U_g$, shown on Fig.~4. The applied negative gate voltage can be
assumed as a wideness of the point contact. This parallel was
explained and used yet in the second section. Increasing $U_g$
means the point contact becomes narrower. Therefore, by changing the
gate voltage we can increase or decrease the conductivity of the
nano-technology point contact, as we act on its wideness. That is
how this simple analogy between electrical potential and wideness
helps us to reduce our problem to the simple model of an usual
Ohm's conductor.

The conductivity is shown in units $2e^2/h$. It's interesting,
that the conductivity is not increasing linearly with the
accretion of the wideness of the point contact, but, as there was
already mentioned, the function has a strange vision, it increases
in some portions, steps or, as we call them - quants, each of
these leading to increasing $\sigma$ with $2e^2/h$. When
$U_g=-2.2$~V the conductivity is zero. This means that there is no
current transferred and that the point contact is stopped. In
other words while $U_g=-2.2$~V the wideness of the gate is zero
and the circuit is ceased, because all the electrons from the
contact are ejected awry due to the electric rejection forces. The
quantization of the conductivity of the unit shown on Fig.~2 in
the mentioned conditions can be detected in the range from $U_g=
-0.3$~V to $U_g = -2.2\;\mathrm{V},$ when the circuit is
disconnected. In this range the conductivity changes altogether 16
times, and each of these portions is equal to $2e^2/h$. The
graphics also shows that the quantization of the conductivity is
not entirely sharp and distinct, the steps are a little bit
rounded. The reason for that indefinite vision of the function is
the resistance of the 2DEG zones, which approximately measures
400~$\Omega.$ Furthermore it is possible to see that the higher is
the conductivity, the smoother is the function. The explanation
for this occurrence in that the 2DEG zones resistance becomes a
bigger part of the whole resistance of the circuit, as the
resistance of the point contact decreases. Now we have the
relation $\sigma/U_g$ entirely explained for $T=0.6$~K. A very
interesting question is to explore the same experiment made in the
conditions of some different temperatures $T.$ The graphic showing
the changes of $\sigma$ as a function of $U_g$ is shown on Fig.~5.
This time there are demonstrated the measured values for different
temperatures: 0.3 K, 0.6 K, 1.6 K, 4.2 K. Obviously the
quantization of the conductivity tends to disappear when
increasing the temperature. Even when $T = 4.2$~K it is really
hard to see it. We have already considered all of the most
interesting and important aspects of the quantization of the
conductivity of the nano-technology point contacts. Now we ought
to mention one very important detail about this problem. We have
already understood where the Ohm's resistance appears in the point
contact, but our work would be incomplete if we overpass the
question where the Ohm's heating becomes; we ought to mention
something about the inconvertible character of the point contact.
The electrons pass through the gate in ballistic regime, so there
is no heat emitted. The heat, actually, is generated when the
electrons transfer from the first Fermi sea to the other through
the one-dimensional channel and they begin to hit into the
surrounding walls. The situation when an electron is coming into a
sea through the point contact reminds the model of the black body
radiation. Exactly as the beam of light, fallen through the narrow
hole into the box cannot went out, exactly the same situation we
have with the electron, fallen into the second 2DEG zone. It
cannot get back to the first Fermi sea through the point-contact.
Therefore the irreversibility of the process of transferring
electrons through nano-technology point contacts appears in the
non-elastic hits, which the electrons bear when coming to the
second, lower level Fermi sea. This process is just the same as
the heating of water, which overflows from one reservoir to
another. In the present work we have solved a contemporary quantum
problem, although the results, necessary for the interpretation of
the experimental data were achieved with elementary methods,
taught in the high-school level of Physics education.
\begin{acknowledgments}
One of the authors (TM) is thankful to T.~Teodosiev  for
invitation this work to be presented at the school in Kazanlak and
to H.~Hristov for the interest and support.
  \end{acknowledgments}

\section{Translation in Slovak language: \'Uvod}
Neust\'ala minia\v{t}uriz\'acia v elektronike n\'uti k tomu, aby
boli kvantov\'e efekty vypo\v{c}\'itavan\'e, po\v{c}as v\'yskumu
nano-\v{s}trukt\'ur. Napr\'iklad v oby\v{c}ajn\'ych meden\'ych
k\'abloch elektr\'ony sa pohybuj\'u presne tak isto,  ako v
objemovom polykri\v{s}t\'alovom kove. Ke\v{d} ale \v{s}irka
vodi\v{c}a je v submikr\'onnom rozsahu, zjavuj\'u sa kvantov\'e
efekty. V t\'ychto mimoriadne tenk\'ych k\'abloch, zvy\v{c}ajne
naz\'yvan\'e nano-k\'able, elektr\'ony sa pohybuj\'u po d\'l\v{z}ke
vodi\v{c}a ako vlna; \v{s}irka k\'abla mus\'i by\v{t} men\v{s}ia ako
100 nm. Mus\'ime vypo\v{c}\'ita\v{t} elektrick\'y pr\'ud,
sp\^osoben\'y rozli\v{c}n\'ymi typmi elektr\'onov\'ych v\'ln,
ktor\'e sa prena\v{s}aj\'u cez nano-k\'abel t\'ym ist\'ym
sp\^osobom, ako sa pren\'a\v{s}a telev\'izny sign\'al cez
koaxi\'alny k\'abel. Ke\v{d} s\'u nano-k\'able dostato\v{c}ne mal\'e
a s\'u prec\'izne vyroben\'e, rozpty\v{l}ovanie elektr\'onov je
dostato\v{c}ne mal\'e a m\^o\v{z}eme o nich uva\v{z}ova\v{t} ako
``letiace'' cez cel\'u d\'l\v{z}ku k\'abla s kon\v{s}tantnou
r\'ychlos\v{t}ou $v$ a momentom $p.$ Tento sp\^osob pohybu sa
naz\'yva balistick\'y, ako vystrelen\'a gu\v{l}ka. \'Uloha pre
vypo\v{c}\'itavanie konduktivity nano-k\'abla, sp\'ajaj\'uc\'eho dva
objemn\'e void\v{c}e, ako je ilustrovan\'e na Fig.~6, bola
vyrie\v{s}en\'a e\'ste pred pribl\'i\v{z}ne p\'ol storo\v{c}\'im
Rolfom Landauerom, ktor\'y vtedy pracoval pre IBM.
\begin{figure}[t]
\centering
\includegraphics[width=0.6\columnwidth]{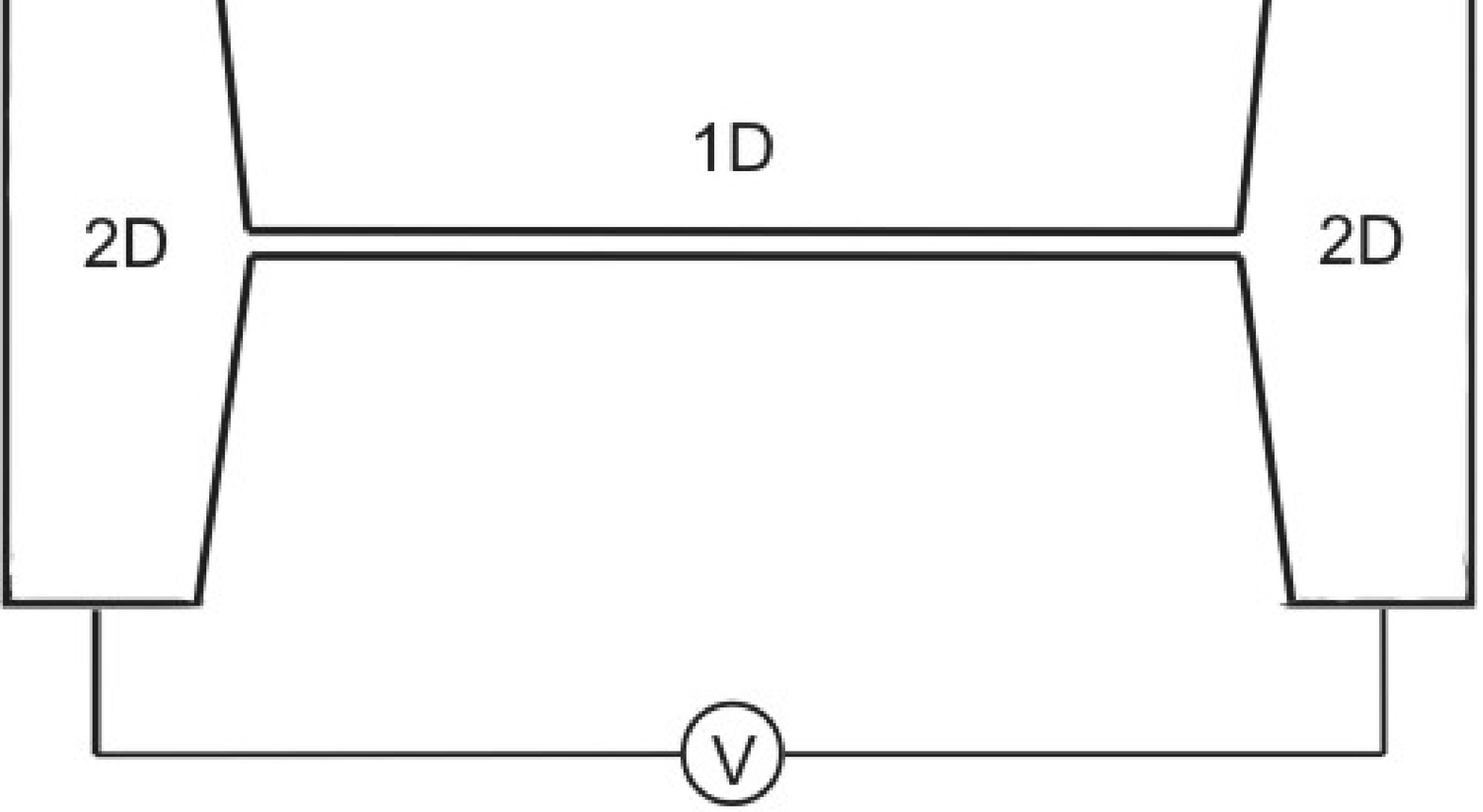}
\caption{Jednodimenzi\'alny (1D) kan\'al sp\'ajaj\'uci dva objemn\'e
dvojdimenzi\'alne (2D) konduktori. Pridan\'e nap\"atie vyvol\'ava
pr\'ud a konduktivita je kvantizovan\'a.
\label{fig:6}}
\end{figure}
Predznamenan\'a od Landauera qantiz\'acia konduktivity bola
demonstrovan\'a ve\v{l}mi prec\'izne od tranzistora s
roz\v{s}tiepen\'ym gejtom \cite{Kouwenhoven,Wees:88,Wees:91},
demonstrovan\'y sch\'ematicky na Fig.~7.
\begin{figure}[t]
\centering
\includegraphics[width=0.6\columnwidth]{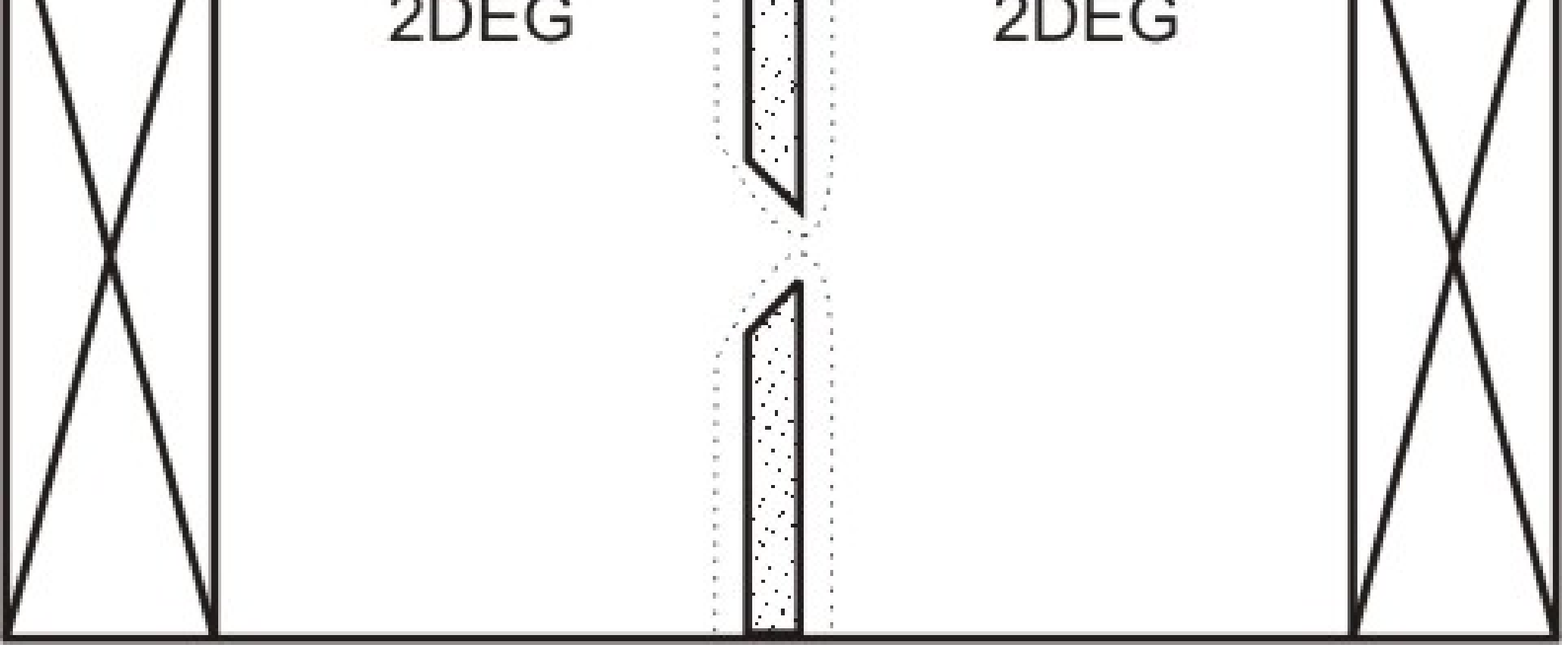}
\caption{Realiz\'acia 1D vodi\v{c}a (bodov\'eho kontaktu) pomocou
tranzistora s roz\v{s}tiepen\'ym gejtom. Zlat\'i gejt odpudzuje
elektr\'ony z dvojdimenzi\'alneho elektronn\'eho plynu (2DEG) v
\'uzkej GaAs vrstve. \'Urovne $\mu_1$ a $\mu_2$ Fermiho mor\'y s\'u
r\^ozli\v{c}n\'e.
\label{fig:7}}
\end{figure}
T\'ato te\'oria sa pou\v{z}\'iva pri analyzovan\'i
submikr\'onn\'ych nano-\v{s}trukt\'ur, ked\v{z}e cel\'a
dne\v{s}n\'a elektronika sa opi\'era na nano-\v{s}trukt\'uru. V
niektor\'ych ved\'ucich elektronick\'ych firm\'ach boli uroben\'e
fundament\'alne sk\'umania vlastnost\'i nano-\v{s}trukt\'ur pri
r\^oznych podmienkach, napr. bolo presk\'uman\'e \'ich chovanie
pri velmi n\'izk\'ych teplot\'ach. Cie\v{l}om nasleduj\'uceho
materi\'alu je op\'isa\v{t} experimenty
\cite{Kouwenhoven,Wees:88,Wees:91} a vyvin\'u\v{t} jednoduch\'y
v\'yvod Landauerovej formuli, s pomocou len element\'arn\'ych
princ\'ipov fyziky, formulovan\'e od Nielsa Bohra. Pre pochopenie
v\'yvodu je potrebn\'e len poznanie Bohrov\'eho atomov\'eho
modelu, ktor\'y sa u\v{z} nieko\v{l}ko desa\v{t}ro\v{c}\'i
vyu\v{c}uje na gymnazi\'alnej \'urovni. S tohto d\^ovodu sa
na\v{s}a pr\'aca mo\v{z}e pou\v{z}\'iva\v{t} od u\v{c}ite\v{l}ov,
ktor\'i by chceli demon\v{s}trova\v{t} svoj\'im \v{z}iakom nov\'y
material z dne\v{s}nej fyziky; niektor\'i \v{z}iaci by mohli
nasleduj\'ucu pr\'acu pochopi\v{t} aj samostatne. Zvy\v{c}ajne pri
sk\'uman\'i nejak\'eho javu  vplyv hranice je pr\'ili\v{s} mal\'y,
aby bol zd\^oraz\v{n}ovan\'y a t\'ym p\'adom sa nepo\v{c}\'ita.
Napr\'iklad tepeln\'a kapacita prstena a cylindra, vyroben\'ych s
toho ist\'eho materi\'alu je prakticky rovnak\'a. V takom
pr\'ipade je pochopite\v{l}n\'e, \v{z}e pou\v{z}ijeme rie\v{s}enie
pomocn\'eho pr\'ikladu pre an\'al\'yzu konduktivity nano-k\'abla.
Tento pomocn\'y pr\'iklad spo\v{c}\'iva v tom, robi\v{t}
v\'po\v{c}ty pre elektr\'ony, cirkuluj\'uce po nano-prstene,
namiesto sk\'umania toku elektr\'onov, ``prelietaj\'ucich'' cez
nano-k\'abel, od negat\'ivneho elektr\'odu k pozit\'ivnemu. My sa
s\'ustrd\'ime na elektr\'ony, cirkuluj\'uce v jednom smere po
nano-prstne s radiusom $r$. Ke\v{d} u\v{z} budeme ma\v{t}
anal\'yzu vz\v{t}ahu elektrick\'eho pr\'udu k nap\"atiu,
``rozre\v{z}eme'' prste\v{n} a vyu\v{z}ijme v\'ysledky pre
konduktivitu nano-technologick\'ych bodov\'ych kontaktov.

\section{Realiz\'acia experimentu}
V nasleduj\'ucich riadkoch op\'i\v{s}eme v kr\'atkosti realiz\'aciu
jednomern\'eho elektr\'onov\'eho vodi\v{c}a pomocou
polovodi\v{c}ov\'ych nano\v{s}trukt\'ur. Zn\'i\v{z}enie
dymenzionality je dosiahnut\'e v dvoch rozdieln\'ych f\'azach, a z
technologick\'eho hladiska je najd\^ole\v{z}itej\v{s}i prechod
vytv\'aranie dvoj-dimenzi\'alneho elektronn\'eho plynu 2DEG. Z tohto
d\^ovodu sa pou\v{z}\'ivaj\'u lamin\'arn\'e polovodi\v{c}ov\'e
\v{s}trukt\'ury. Tenk\'a v\'rstva GaAs sa nan\'a\v{s}a do AlGaAs.
Bl\'izko tejto v\'rstvy sa objemn\'y AlGaAs j\'onizuje v tenkej
v\'rstve so Si. Elektr\'on vybieha z Si donorov a pad\'a do
potenci\'alnej jamy GaAs. V tejto situ\'acii \'ich vlnov\'e funkcie
s\'u jednoducho stoj\'ace vlny, ako tie, \v{c}o sa vytv\'araj\'u pri
chven\'i struny husli\v{c}iek. V plo\v{s}ine prostredia sa ale
elektr\'ony pohybuj\'u ako voln\'e dvoj-dymenzi\'aln\'e \v{c}astice
a ich vlnov\'e funkcie s\'u plosk\'e vlny. In\'imi slovami,
m\^o\v{z}eme po\v{c}\'ita\v{t}, \v{z}e elektr\'ony sa
vzn\'a\v{s}aj\'u ako \v{c}ajky nad morskou h\v{l}adinou. \'Ich
rozpt\'ylenie na vzdialen\'ych j\'onizovan\'ych Si donoroch je
ve\v{l}mi mal\'e a preto konduktivita 2DEG-u je velmi vysok\'a.
Hustota tohto elektronn\'eho plinu je ve\v{l}k\'a, a mo\v{z}eme
uva\v{z}ova\v{t}, \v{z}e m\'ame dvoj-dymenzi\'aln\'i kov pri
n\'izk\'ych teplot\'ach. Tento kov je ve\v{l}mi dobre izolovan\'y ot
semikonduktora, v ktorom je umiestnen\'y. Pre t\'uto \v{s}trukt\'uru
Pauliho z\'abrana \'u\v{c}inkuje a v\v{s}etky elektr\'ony
vyp\'l\v{n}aj\'u v\v{s}etky elektr\'onn\'e \'urovne po nejak\'u
maxim\'alnu mo\v{z}n\'u energiu $E_\mathrm{max},$ zvanou Fermiho
energia, ako je schematicky uk\'azan\'e na Fig.~8.
\begin{figure}[t]
\centering
\includegraphics[width=0.6\columnwidth]{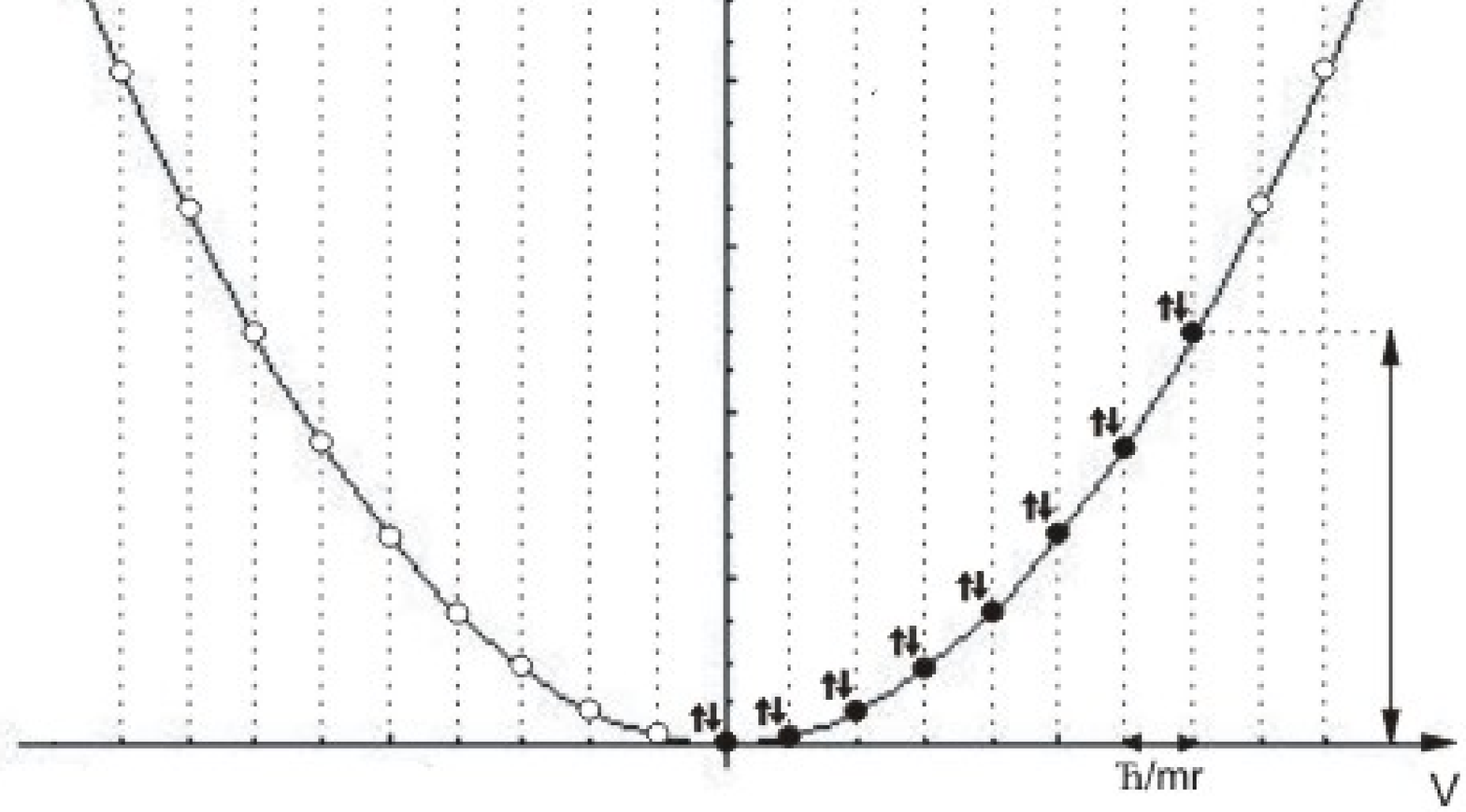}
\caption{Kvantiz\'acia energie elektr\'onov, cirkulujuc\'ich po
okruhu s r\'adiusom $r.$ Podla Bohrov\'ych z\'akonov uhlov\'y moment
elektr\'onu je kvantizovan\'y. S toho je sp\^osobena kvantiz\'acia
aj r\'ychlosti aj energie. Pr\'azdn\'e kr\'u\v{z}ky
demon\v{s}truj\'u pr\'azdn\'e elektr\'onn\'e orbity. Pln\'e
kr\'u\v{z}ky ukazuj\'u dvojito zaujat\'e \'urovne. Elektr\'ony
sprava maj\'u maxim\'aln\'u energiu.
\label{fig:8}}
\end{figure}

Predt\'ym spomenut\'e  zap\'l\v{n}anie lektr\'onnych poz\'ici\'i je
analogick\'e zap\'l\v{n}aniu pr\'azdnych medzier na zemskom povrchu
vodou po \'urove\v{n} morskej hladiny; z tohto d\^ovdu sa \v{c}asto
pou\v{z}\'iva fr\'aza Fermiho more. V na\v{s}om pr\'ipade more je
dvojdymenzi\'alne. Technologick\'y probl\'em je vyrobi\v{t}
jednomern\'y kan\'al medzi dvomi podobn\'ymi morami. Na dosiahnutie
tohto cie\v{l}a je pou\v{z}it\'y roz\v{s}tiepen\'y gejt z dvoch
zlat\'ych elektr\'od, vyparen\'y na polovodi\v{c}ov\'u
hetero\v{s}trukt\'uru. T\'ato met\'oda na v\'yrobu tranzistora s
roz\v{s}tiepen\'ym gejtom sa pou\v{z}\'iva, a je demon\v{s}trovan\'a
na Fig.~7. S\'orce a stok maj\'u ohmov\'e spojenie s 2DEG. Ke\v{d}
gejtu prid\'ame silne negat\'ivn\'y potenci\'al, elektr\'ony pod
n\'im zmizn\'u, vytla\v{c}en\'e nabok coulombov\'ymi odpudzuj\'ucimi
silami gejtu.
\begin{figure}[t]
\centering
\includegraphics[width=0.6\columnwidth]{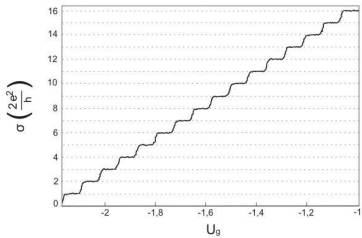}
\caption{Konduktivita $\sigma$ tranzistora oproti gejtovej
volt\'a\v{z}e. Zv\'y\v{s}enie nap\"atia otv\'ara nov\'e
elektr\'onn\'e kan\'aly. Ka\v{z}d\'y elektr\'onn\'y kan\'al d\'ava o
jeden kondukt\'ivn\'y kvantum $2e^2/h$ celej konduktivite, a
v\v{d}aka tomu sa vytv\'ara v\'y\v{s}ka schodou.
\label{fig:9}}
\end{figure}
T\'ymto sp\^osobom sa spojenie medzi dvoma morami preru\v{s}\'i a
tranzistor sa upch\'a. Ke\v{d} silu odpudzovania zmen\v{s}\'ime,
zmenen\'im gejtov\'eho nap\"atia kontakt medzi Fermi morami sa znovu
vytv\'ara a elektr\'ony prechadzaj\'u z jedn\'eho mora do druh\'eho
cez \'uzky jednodymenzi\'alny kan\'al, vytvoren\'y pod
roz\v{s}tiepen\'ym gejtom. V tejto f\'aze transf\'eru elektr\'ony
maj\'u skoro jednosmern\'y sp\^osob pohybu - ako vlny vo vlnovode.
Oni sa nerozptluju, ale letia ako n\'aboje. Z tohto d\^ovodu
vol\'ame tento typ pohybu balistick\'y re\v{z}\'im. Vlnov\'e funkcie
s\'u plosk\'e vlny, rozsahuj\'uce sa pozd\'l\v{z} kan\'ala, v
perpendikul\'arnej rovine s\'u to stojat\'e vlny. Zaplnenie
elktr\'onn\'ych \'urovn\'i z\'ale\v{z}\'i od potenci\'alu, tak\v{z}e
gejtov\'a volt\'a\v{z} ur\v{c}uje po\v{c}et otvoren\'ych
elektr\'onn\'ych kan\'alov. Ka\v{z}d\'y z t\'ychto kan\'alov d\'ava
jeden kvantum konduktivity k celej konduktivity
nano-technologick\'eho bodov\'eho kontaktu. Teraz u\v{z}
m\^o\v{z}eme pochopi\v{t} ``schodovit\'y'' charakter vzh\v{l}adu
rel\'acie medzi konduktivitou a gejtov\'ym nap\"at\'im pri
n\'izk\'ych teplot\'ach, zaobrazenou na Fig.~9. Ka\v{z}d\'y
``schod'' znamen\'a, \v{z}e je otvoren\'y \v{d}al\v{s}\'i
jednomern\'y kan\'al medzi Fermiho morami. Zv\'y\v{s}enie teploty
prid\'ava ``schodom'' hladkej\v{s}\'i v\'yzor. Postupne sa stan\'u
neodhaliteln\'ymi (Fig.~10); ochl\'adzovanie n\'am prin\'a\v{s}a
zd\^oraznenej\v{s}ie schodito vyzeraj\'ucu figuru. V\'y\v{s}ka
schodou m\^o\v{z}e by\v{t} vypo\v{c}\'itan\'a element\'arnymi
poznatkami. Tieto kalkul\'acie s\'u realizovan\'e v nasleduj\'ucej
sekcii tohto textu.
\begin{figure}[t]
\centering
\includegraphics[width=0.6\columnwidth]{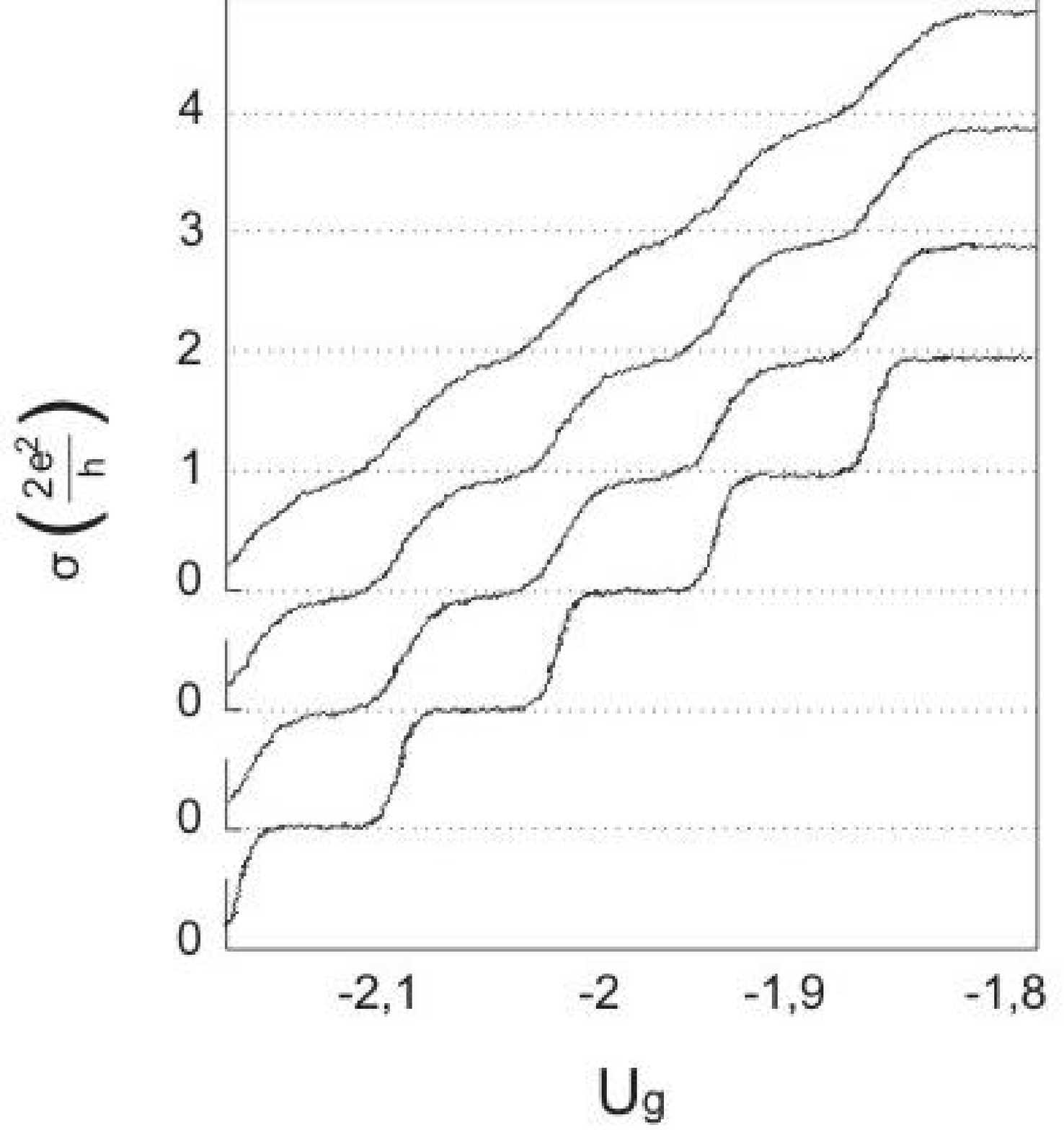}
\caption{Konduktivita $\sigma$ oproti gejtov\'emu volt\'a\v{z}u
$U_g$ pri r\^ozn\'ych teplot\'ach. Zv\'y\v{s}enie teploty rob\'y
schody kvantiz\'acie hlad\v{s}ie. Pri vysok\'ych teplot\'ach
kvantiz\'acia konduktivity mizne.
\label{fig:10}}
\end{figure}

 Aby sme boli prestn\'i by sme museli po\v{c}\'ita\v{t} aj
 elektrick\'e pr\'udy, prebiehaj\'uce v opa\v{c}nom smere, vtedy,
 ke\v{d} $U$ je mal\'y rozdiel medzi rovinami Fermiho mor\'y;
 nez\'avisle od toho v\'ysledky zost\'avaj\'u tie ist\'e. To je v\v{s}etko,
 \v{c}o povieme o realiz\'acii experimentu. V nasleduj\'ucej sekcii bude
 predstaven\'y v\'yvod na Landauerov\'u formulu, opisuj\'ucu  kvantiz\'aciu
 konduktivity. \v{D}alej charakterizujeme vplyv zmeny teploty na
 konduktivitu nano-technologick\'eho bodov\'eho
 kontaktu.

\section{V\'yvod formule}
U\v{z} sme spomenuli, \v{z}e budeme analyzova\v{t} myslen\'y
experiment s elektr\'onmi, cirkuluj\'ucimi po okruhu s r\'adiusom
$r$, ako v \'atomovom modele Bohra. Teneto experiment n\'am
pom\^o\v{z}e pri rie\v{s}en\'i na\v{s}ej \'ulohy. Ke\v{d} sa
elektr\'on to\v{c}\'i r\'ychlos\v{t}ou $v$ peri\'oda jeho
ot\'a\v{c}ky je $2\pi r/v$ (\v{c}as sa vyjadruje ako prejden\'a
cesta, rozdelen\'a na r\'ychlos\v{t}). Stredn\'y elektrick\'y pr\'ud
$I_n$ je proste n\'aboj elektr\'onu $e$ rozdelen\'y na peri\'odu
(defin\'icia pre elektrick\'y pr\'ud je prebehnut\'i n\'aboj za
ur\v{c}it\'y \v{c}as)
\begin{equation}
\label{current} I_n=\frac{ev_n}{2\pi r}.
\end{equation}
Index n znamen\'a, \v{z}e re\v{s}pektujeme kvantov\'e vlastnosti
elektr\'onu. Cel\'e \v{c}\'islo $n$ ukazuje ko\v{l}ko kr\'at
d\'l\v{z}ka vlny elektr\'onu $\lambda_e=2\pi\hbar/p$ sa nach\'adza v
perimetri kruhu $2\pi r=n \lambda_e$. Pod\v{l}a Bohrovho z\'akona
Uhlov\'y moment je kvantizovan\'y
\be mvr=n\hbar, \ee
kde $\hbar=1.055\times 10^{-34}\,\mathrm{Js}$ je Plankov\'a
kon\v{s}tanta a cel\'a \v{c}islica $n$ m\'a n\'azov magnetov\'e
kvantov\'e \v{c}\'islo v \'atomovej fyzike, $m$ je efekt\'ivna
hmotnos\v{t} elektr\'onu v kri\v{s}talickej \v{s}trukt\'ure. Z
Bohrovho z\'akona m\^o\v{z}eme vypo\v{c}\'ita\v{t} r\'ychlos\v{t}:
\be v_{n}=\frac{\hbar}{mr}n .\ee
V\'ysledky m\^o\v{z}eme pou\v{z}i\v{t} vo formule o eletrickom
pr\'ede Eq.~(\ref{current}).
\be I_n=\frac{e\hbar}{2\pi mr^2}n. \label{1 electron current } \ee
Kvantiz\'acia r\'ychlosti sp\^osobuje kvantiz\'aciu kinetickej
energie tie\v{z}
\be E_n=\frac{1}{2}mv_n^2=\frac{1}{2}\frac{\hbar^2}{mr^2}n^2. \ee
T\'ato formula opisuje energetick\'e spektrum elektr\'onu,
To\v{c}iacim sa po kruhu s fixovan\'ym r\'adiusom, graficky
uk\'azanom na Fig.~8. Vypo\v{c}\'itavanie cel\'eho pr\'udu,
sp\^osoben\'eho v\v{s}etk\'ymi elektr\'onmi, kr\'u\v{z}iacimi v
jednom smere $v>0,$ u\v{z} nie je pobl\'em, pova\v{z}uj\'uc, \v{z}e
v\v{s}etky elektr\'onne \'urovne v interv\'ale od $E=0$ da
$E=E_\mathrm{max}$ s\'u zaplnen\'e presne dvoma elektr\'onmi
ka\v{z}d\'a (zo spinom ``hore'' a ``dolu''). Pod\v{l}a Pauliho
z\'abrane ka\v{z}d\'a elektr\'onna \'urovev{n} m\^o\v{z}e by\v{t}
zaplnen\'a len jedn\'ym elektr\'onom s dan\'ym spinom a z formule o
spektre je jasn\'e, \v{z}e maxim\'alna energia $E_\mathrm{max}$
m\^o\v{z}e by\v{t} parametrizovan\'a s jednou ve\v{l}kou celou
\v{c}islicou $N$ alebo s \v{c}lenom, maj\'ucim rozmernos\v{t}
elektrick\'eho potenci\'alu $U$
\be E_\mathrm{max}=\frac{1}{2} \frac{\hbar^2}{mr^2} N^2 =
eU.\label{Fermi}\ee
Cel\'y pr\'ud je suma zo v\v{s}etk\'ych elektr\'onnych pr\'udov
\be I= 2 \sum^{N}_{n=0} I_n.\label{total current} \ee
S in\'ymi slovami, mus\'i sa sumova\v{t} cez v\v{s}etky elektr\'onne
\'urovne. M\^o\v{z}eme pou\v{z}i\v{t} formulu o aritmetickej
progr\'esii pre dostato\v{c}ne ve\v{l}k\'e \v{c}\'islo $N\gg1$
\be \sum_{n=1}^N n=1+2+3+\dots+N=\frac{N(N+1)}{2}\approx
\frac{N^2}{2}\gg1. \ee
Pou\v{z}ijeme t\'uto formulu pre kalkul\'aciu pr\'udov
Eq.~(\ref{total current}), vyjadren\'imi cez Eq.~(\ref{1 electron
current })
\be I= 2 \sum^{N}_{n=0} I_n
 = 2\sum^{N}_{n=0}\frac{e\hbar}{2\pi mr^2}n
 = \frac{2e\hbar}{2\pi mr^2} \frac{N(N+1)}{2}.
\ee
Cel\'a \v{c}islica $N$ je ove\v{l}a ve\v{c}\v{s}ia ako 1, tak\v{z}e
s dostato\v{c}nou presnos\v{t}ou m\^o\v{z}eme prija\v{t}, \v{z}e
$N(1+1/N)\approx N$ a dostaneme
\be I= 2\frac{e}{2\pi\hbar} \left(\frac{\hbar^2N^2}{2mr^2}\right).
\ee
V\'yraz v z\'atvork\'ach je maxim\'alna mo\v{z}n\'a energia
elektr\'onov, spomenut\'ych v Eq.~(\ref{Fermi}). Teraz m\'ame
mo\v{z}nos\v{t} vyjadri\v{t} pr\'ud pomocou elektrick\'eho
potenci\'alu
\be I=\frac{2e^2}{2\pi\hbar}U=\frac{2e^2}{h}U=\sigma U, \ee
Pod\v{l}a trad\'icie sme pou\v{z}ili star\'u Plankov\'u
kon\v{s}tantu $h=2\pi\hbar.$ Pod\v{l}a defin\'icie, konduktivita
$\sigma$ sa rov\'na pr\'udu, rozdelen\'emu nap\"atiu $\sigma=I/U$. V
celej kalkul\'acii sme po\v{c}\'itali s t\'ym, \v{z}e re\v{z}\'im
elektr\'onov, pren\'a\v{s}an\'ych cez prestenovidny vodi\v{c}, je
balistick\'y. Pri tomto re\v{z}ime elektr\'onne rozpt\'ilnie je
zanedbate\v{l}n\'e a preto d\'l\v{z}ka vlny elektr\'onov mus\'i
by\v{t} men\v{s}ia ako stredn\'y voln\'y priebeh elektr\'onu.
V\'ysledok, ktory sme dostali je mo\v{z}n\'e pou\v{z}i\v{t} pre
dostato\v{c}ne kr\'atke nano-k\'able a taktie\v{z} bodov\'e
kontakty, kde sa prechod uskuto\v{c}\v{n}uje pomocou
jednodimenzi\'aln\'eho pohybu. Teraz u\v{z} m\'ame Landauerov\'u
formulu, opisuj\'ucu kvantiz\'aciu konduktivity
jednodimenzi\'aln\'eho void\v{c}a.
\be \sigma_0=\frac{2e^2}{h}
 = \frac{1}{12906\;\Omega}
 = 77.5\,\mu\mathrm{Sm}.
\ee
V re\'alnom nano-technologickom bodovom kontakte je konduktivita
dosiahnut\'a pomocou ve\v{l}k\'eho mno\v{z}stva podobn\'ych
jednodimenzi\'aln\'ych kan\'alov a pre v\'ypo\v{c}et celej
konduktivity mus\'ime n\'asobi\v{t} kvantum konduktivity  $\sigma_0$
s po\v{c}tom otvoren\'ych jednodimenzi\'aln\'ych kan\'alov $K$
\be \sigma = K\sigma_0= K \frac{2e^2}{h}.
\label{p2hia}
\ee

\section{Vplyv teploty na konduktivitu}
Presk\'umajme v\'ysledky, dosiahnut\'e pri realiz\'acii experimentu.
V na\v{s}ej pr\'aci sme pou\v{z}ili tranzistor, zobrazen\'i na
Fig.~7. Teplota, pri ktorej boli uskuto\v{c}\v{n}en\'e merania je
bl\'izka k absolutnej $T = 0.6$~K. Dost\'avaj\'uc inform\'aciu o
volt\'a\v{z}i z pr\'iborov m\^o\v{z}eme \v{l}ahko
vypo\v{c}\'ita\v{t} konduktivitu pre r\^ozne ve\v{l}kosti
negat\'ivneho gejtov\'eho volt\'a\v{z}u $U_g$, prilo\v{z}en\'emu k
bodov\'emu kontaktu. S t\'ymto negat\'ivnym volta\v{z}om sa
m\^o\v{z}e urobi\v{t} jednoduch\'a anal\'ogia, ktor\'a n\'am
pom\^o\v{z}e pochopi\v{t} grafick\'u s\'uvislos\v{t} medzi $\sigma$
a $U_g$, uk\'azan\'u na Fig.~9. Pridan\'y gejtov\'y potencial
m\^o\v{z}e by\v{t} prijat\'y ako \v{s}\'irka gejtu. T\'ato
anal\'ogia u\v{z} bola vyu\v{z}it\'a v druhej sekcii.
Zve\v{c}\v{s}enie $U_g$ Znamen\'a, \v{z}e sa bodov\'y kontakt stane
uz\v{s}\'im. To znamen\'a, \v{z}e menenie negat\'ivn\'eho nap\"atia
na gejte mo\v{z}e sp\^osobi\v{t} sp\'ad, alebo zvy\v{s}enie
konduktivity nano-technologick\'eho bodov\'eho kontaktu, zmenen\'im
jeho \v{s}\'irky. Takto t\'ato jednoduch\'a anal\'ogia medzi
potenci\'alom a \v{s}\'irkou n\'am umo\v{z}nila redukovanie
na\v{s}ej \'ulohy k prost\'emu modelu oby\v{c}ajn\'emu Ohmov\'emu
vodi\v{c}u.

Konduktivita je predstaven\'a v jednotk\'ach $2e^2/h$. Zaujmav\'e
je, \v{z}e sa konduktivita nezvy\v{s}uje line\'arne s line\'arn\'ym
zve\v{c}\v{s}en\'im  \v{s}\'irky bodov\'eho kontaktu, ale, ako sme
u\v{z} op\'isali, t\'ato funkcia m\'a zvl\'a\v{s}tny vzh\v{l}ad;
konduktivita sa zve\v{c}\v{s}uje v ur\v{c}it\'ych porci\'ach,
d\'ozach, naz\'yvan\'e kvanty. Ka\v{z}d\'y z n\'ich zv\'y\v{s}uje
konduktivitu $\sigma$ o $2e^2/h$. Ke\v{d} $U_g=-2.2$~V konduktivita
je nulov\'a. S toho vypl\'yva. \v{z}e \v{z}iadn\'y pr\'ud
neprech\'adza a bodov\'y kontakt je uzatvoren\'y. In\'ymi slovami,
ke\v{d} $U_g=-2.2$~V \v{s}\'irka gejtu je nulov\'a a re\v{t}az je
preru\v{s}en\'a, preto\v{z}e v\v{s}etky elektr\'ony z kontaktu s\'u
vytla\v{c}en\'e nabok ot elektrick\'ych odpudzujuc\'ich s\'il.
Kvantiz\'acia konduktivity pr\'istroja, uk\'azanom na Fig.~7 v
spomenut\'ych podmienkach m\'o\v{z}e by\v{t} detektovan\'a v
hraniciach d $U_g= -0.3$~V do $U_g= -2.2\;\mathrm{V},$ ke\v{d} je
re\v{t}az preru\v{s}en\'a. v t\'ychto hraniciach sa konduktivita
men\'i spolu 16 kr\'at, a ka\v{z}d\'a z porci\'i sa rovn\'a
$2e^2/h$. Grafika taktie\v{z} ukazuje, \v{z}e kvantiz\'acia
konduktivity nie je celkom prudk\'a a presn\'a, ``schody' s\'u
trochu obl\'e. Pr\'i\v{c}ina tohto neide\'aln\'eho vzh\v{l}adu
funkcie je elektrick\'y odpor 2DEG z\'on, ktor\'y je pribli\v{z}ne
400~$\Omega.$ \v{D}alej m\^o\v{z}eme spozorova\v{t}, \v{z}e
\v{c}\'im je vy\v{s}\v{s}ia konduktivita, t\'ym je funkcia
oblej\v{s}ia. Vysvetlenie pre tento jav n\'ajdeme v opore 2DEG
z\'on, ktor\'i sa stane ve\v{c}\v{s}iou \v{c}iastkou cel\'eho odporu
re\v{t}aze, k\^oli zn\'i\v{z}eniu rezistencie bodov\'eho kontaktu.
Teraz u\v{z} m\'ame pomer $\sigma/U_g$ vyjasnen\'y pre $T=0.6$~K.
Ve\v{l}mi zaujimav\'a \'uloha je presk\'uma\v{t} ten ist\'y
experiment, preveden\'y pri r\^oznych teplot\'ach $T.$ Grafika,
zobrazuj\'uca zmeny $\sigma$ ako funkcia $U_g$ je uk\'azan\'a na
Fig.~10. Tento raz s\'u tam zobrazen\'e zmeran\'e hodnoty  pre
r\^ozne teploty: 0.3 K, 0.6 K, 1.6 K, 4.2 K. Zrejme kvantiz\'acia
konduktivity postupne mizne so zv\'y\v{s}en\'im teploty. E\v{s}te
pri $T = 4.2$~K je u\v{z} len \v{t}a\v{z}ko rozl\'i\v{s}iteln\'a.
U\v{z} sme skon\v{c}ili s op\'isovan\'im ve\v{c}\v{s}iny
najd\^ole\v{z}itej\v{s}\'ich a najzaujmavej\v{s}\'ich aspekt\'ov
kvantiz\'acie konduktivity nano-technologick\'eho bodov\'eho
kontaktu. Teraz by sme chceli op\'isa\v{t} jeden ve\v{l}mi
d\^ole\v{z}it\'y detajl, t\'ykajuc\'i sa tejto pr\'ace. U\v{z} sme
op\'isali kde vznik\'a Ohmova rezistencia v na\v{s}om pr\'ibore, ale
na\v{s}\'a pr\'aca by bola neupln\'a, keby sme vynechali ot\'azku
kde presne sa Ohmov\'e zahrievanie prejavuje; na tomto mieste by sme
chceli spomen\'u\v{t} nie\v{c}o o neobratitelnosti bodov\'eho
kontaktu, pri zv\'y\v{s}en\'i entropie. Elektr\'ony prech\'adzaj\'u
cez gejt v balistickom re\v{z}\'ime, tak\v{z}e tu sa teplota
nevylu\v{c}uje. Teplota sa vlastne prejavuje pri prechode
elektr\'onov z jedn\'eho Fermiho mora do druh\'eho cez
jednodimenzi\'alny kan\'al a za\v{c}\'inaj\'u sa udiera\v{t} do
okolit\'ych stien. Situ\'acia, pri ktorej jeden elektr\'on vch\'adza
do mora cez bodov\'y kontakt pripom\'ina model vysielania uplne
\v{c}iern\'eho telesa. Presne tak isto ako l\'u\v{c} svetla,
padaj\'uci cez \'uzky otvor do \v{s}katule ju u\v{z} nem\^o\v{z}e
opusti\v{t}, aj elektr\'on, pre\v{s}iel do druhej 2DEG z\'ony ju
u\v{z} nem\^o\v{z}e opusti\v{t} cez jednodimenzi\'aln\'y kan\'al.
Neobratitelnos\v{t} procesu prenosu elektr\'onov cez
nano-technologick\'y kontakt sa prejavuje pr\'ave v neelastick\'ych
\'uderoch, ktor\'e elektr\'ony utrpia ke\v{d} prejdu do druh\'eho,
ni\v{z}\v{s}ieho Fermiho mora. Tento proces je naplno identick\'y so
zahrievan\'im vody, ktor\'a sa prelieva z jedn\'eho rezervo\'aru do
druh\'eho. V na\v{s}ej pr\'aci sme vyrie\v{s}ili jedn\'u zaujimav\'u
\'ulohu z aktu\'alnej fyziky, aj ke\v{d} v\'ysledky, potrebn\'e pre
anal\'yzu experiment\'aln\'ych d\'at sme dosiahli s element\'arnimi
met\'odami, vyu\v{c}ovan\'ymi na stredno\v{s}kolskej \'urovni
vyu\v{c}ovania z fyziky.
\begin{acknowledgments}
Jeden z autorov (TM) \v{d}akuje T.~Teodosievovi  za pozvanie pre
predn\'a\v{s}anie tohto textu v \v{s}kole v Kazanlaku a H.~Hristovi
za z\'aujem a podporu.
\end{acknowledgments}

\begin{references}
\bibitem{Wees:88} B.J.~van~Wees, H.~van~Houten, C.W.J.~Beenaker, J.G.~Williamson,
  L.P.~ Kouwenhoven, D.~van~der~Marel, and C.T.~Foxon,
  Phys. Rev. Lett. \textbf{60}, 848 (1988).

\bibitem{Wees:91} B.J.~van~Wees, L.P.~Kouwenhoven, E.M.M.~Willems,
  C.J.P.M.~Harmans, J.E.~Mooij, H.~van~Houten, J.G.~Williamson, and C.T.~Foxon,
  Phys. Rev. B \textbf{43}, 12431 (1991).

\bibitem{Kouwenhoven}L.P.~Kouwenhoven,
  ``Transport of Electron-Waves and Single-Charges in Semiconductor Nanostructures'',
  Ph.D Thesis Technische Universiteit Delft (1992); Chap.~I.
\end{references}
\end{document}